\documentclass[pra,twocolumn,aps,amsmath,showkeys,showpacs,floatfix,nofootinbib]{revtex4-1}
\usepackage{graphicx}
\usepackage{dcolumn}
\usepackage{bm}
\usepackage{color}

\def\beq{\begin{equation}}
\def\eeq{\end{equation}}
\def\beqa{\begin{eqnarray}}
\def\eeqa{\end{eqnarray}}
\begin{document}


\title{Fast generation of spin-squeezed 
states in bosonic Josephson junctions}

\author{B. Juli\'a-D\'\i az$^1$, E. Torrontegui$^2$, 
J. Martorell$^3$, J. G. Muga$^{2,4}$, and A. Polls$^3$}

\affiliation{$^1$ICFO-Institut de Ci\`encies Fot\`oniques, 
Parc Mediterrani de la Tecnologia, 08860 Barcelona, Spain}
\affiliation{$^2$ Departamento de Qu\'\i mica-F\'\i sica, 
UPV-EHU, Apartado 644, 48080 Bilbao, Spain}
\affiliation{$^3$Departament d'Estructura i Constituents de la Mat\`eria, 
Facultat de F\'isica, U. Barcelona, 08028 Barcelona, Spain}
\affiliation{$^4$ Department of Physics, Shanghai University, 
200444 Shanghai, People’s Republic of China}
\begin{abstract}

We describe methods for the fast production of highly 
coherent-spin-squeezed many-body states in bosonic 
Josephson junctions (BJJs). We start from the known 
mapping of the two-site Bose-Hubbard (BH) Hamiltonian 
to that of a single effective particle evolving according 
to a Schr\"odinger-like equation in Fock space. Since, 
for repulsive interactions, the effective potential in 
Fock space is nearly parabolic, we extend recently 
derived protocols for shortcuts to adiabatic evolution 
in harmonic potentials to the many-body BH Hamiltonian. 
A comparison with current experiments shows 
that our methods allow for an important reduction in 
the preparation times of highly squeezed spin states.
\end{abstract}

\maketitle

\section{Introduction}

Recent experimental results from the groups of M. 
Oberthaler and P. Treutlein have provided compelling 
evidence of the generation of spin-squeezed states 
in bosonic Josephson junctions~\cite{esteve08,gross10,riedel10,zib10}. 
Two different setups with ultracold $^{87}$Rb bosons 
have been explored: a) a cloud trapped in an 
external double-well potential~\cite{Albiez05,GO07,esteve08} 
(external BJJ) and b) a cloud of atoms in two 
different hyperfine states trapped in a single 
harmonic potential with a linear coupling between 
the two internal states~\cite{gross10,riedel10,zib10} (internal BJJ). 
In both the internal and external Josephson junctions, 
the experimental setups were able to reproduce to a 
large extent the well-known two-site Bose-Hubbard (BH)
model~\cite{Mil97,Leggett01,GO07}, a Lipkin-Meshkov-Glick-like 
(LMG) Hamiltonian~\cite{lipkin}. In this way, the 
experiments confirmed the predicted 
existence of strongly squeezed spin states~\cite{kita} 
and, correspondingly, of large sets of entangled 
pseudo-spins~\cite{Sorensen2001,korb05}. 

Simultaneously, and due to the need to control and 
implement quantum resources, there has been an 
increasing interest in developing fast protocols to 
shortcut adiabatic following~\cite{muga1,muga2}. The 
key purpose is to engineer procedures to drive, in 
a finite time, a system from some initial state to 
a final state that could be reached with an adiabatic, 
slow process. The different protocols may in addition 
aim at minimizing the transient energy 
excitations, reducing the sensitivity to noise, 
or minimizing other variables of interest~\cite{muga4}. 
It should be noted that, in general, the initial and 
final states are not necessarily required to be the 
ground states (GSs) of the system. In this work, 
however, we concentrate on transitions between ground 
states corresponding to different values of the model 
parameters which can be controlled externally. 
The proposed method is designed so that the desired 
final state is produced as a stationary eigenstate of 
the Hamiltonian, with no need to freeze the 
dynamics. Analytical formulas to perform this type of 
processes exist for the harmonic oscillator~\cite{muga1}.

In this article we demonstrate that these 
methods can also be employed to produce
the highly-correlated many-body quantum states 
described by the BH Hamiltonian, such as squeezed 
states, broadening 
the current state-of-the-art which has so far 
dealt only with condensed cold gases~\cite{schaff10}. 
To do so, we benefit from the explicit mapping, 
in the large $N$ limit, between the two-site Bose-Hubbard 
model and an effective single particle system 
described by a Schr\"odinger-like 
equation~\cite{ST08,ours10-2,oursober}\footnote{In 
the context of the LMG model, this connection was 
established earlier~\cite{lmg1,lmg2}, albeit with 
no known relation at the time with ultracold atom 
physics.}. As will be shown, this connection 
allows us to use methods which were originally developed for 
single-particle dynamics, to shortcut the adiabatic 
following in a many-body problem. As a figure of merit, 
the methods proposed would allow to prepare highly 
number-squeezed states in roughly half the time needed 
in the experimental setup of Ref.~\cite{riedel10}.

Our proposal presents important differences with 
respect to the recent work of Ref.~\cite{lapert12}. We 
target the formation of spin-squeezed states 
while ~\cite{lapert12} uses optimal control theory 
(OCT) to produce cat-like states. In our method, the 
final state of the system is the ground state, with 
no need to stop or freeze the dynamics, while 
in ~\cite{lapert12} a stoppage is required once 
the desired state is reached. Also, as detailed 
below, our method to shortcut the adiabatic 
evolution is very robust, only requiring high 
precision control during the initial and 
final times.

The article is organized as follows. In Sec.~\ref{sec1} we 
present the Bose-Hubbard model, and introduce the semiclassical 
1/N approximate model. In Sec.~\ref{sec2} we present the 
methods to shortcut the adiabatic evolution and adapt them to 
our specific problem. In Sec.~\ref{sec3} we discuss the obtained 
results. Finally, Sec.~\ref{sec4} presents our conclusions. 

\section{Description of the model}
\label{sec1}

We assume that in the BJJ, the system of ultra-cold 
bosons is well modeled by the Bose-Hubbard 
Hamiltonian $\hbar {\cal H}_{\rm BH}$ 
\beq
{\cal H}_{\rm BH}= -2J \hat{J}_x + U \hat{J}_z^2  \,,
\label{eq:bh}
\eeq
where the pseudo-angular momentum operator 
$\hat{{\bf J}}\equiv \{\hat{J}_x,\hat{J}_y,\hat{J}_z\}$ 
is defined as,
\beqa
{\hat J}_x &=& {1\over2} (\hat{a}_1^{\dag} \hat{a}_2 + \hat{a}_2^{\dag} \hat{a}_1)\,,
\nonumber\\ 
{\hat J}_y &=& {1\over2i}(\hat{a}_1^{\dag} \hat{a}_2 - \hat{a}_2^{\dag} \hat{a}_1) \,,
\\
{\hat J}_z &=& {1\over2} (\hat{a}_1^{\dag} \hat{a}_1 - \hat{a}_2^{\dag} \hat{a}_2)\,,
\nonumber
\eeqa 
where $\hat{a}_j^{\dag}$ creates a boson in site $j$,  and 
$[\hat{a}_i,\hat{a}_j^{\dag}] =\delta_{i,j}$. 
$J$ is the hopping strength, taken positive, 
and $U$ is the non-linear coupling strength 
proportional to the atom-atom $s$-wave scattering 
length\footnote{In the internal BJJ, $U$ is 
proportional to $a_{1,1}+a_{2,2}-2 a_{1,2}$, with 
$a_{1,1}$ and $a_{2,2}$ the intra-species scattering 
lengths and $a_{1,2}$ the inter-species 
one~\cite{zib10}.}. In this work we consider repulsive 
interactions, $U>0$. For internal BJJ, the inter-species 
$s$-wave scattering length in $^{87}$Rb atoms can 
be varied by applying an external magnetic field 
thanks to a well characterized Feshbach resonance 
at $B =9.1$ G, as done in Ref.~\cite{zib10} for the setup 
that we are considering, thus permitting precise control 
over the $U$. In this work, we use a time 
dependent $U(t)$, keeping $J$ and $N$ fixed during 
the time evolution. This has been experimentally 
achieved already in an atom-chip experiment~\cite{riedel10}. 
In the experiments of the Heidelberg group one would 
need to track the resonance of the linear coupling (responsible 
for the hopping term in the Hamiltonian) which is 
altered due to the second order Zeeman shift when 
$B$ is varied. Albeit technically challenging, this 
is within reach of the current experimental 
setups~\cite{zibpriv}.

The time dependent Schr\"odinger equation (TDSE) is 
written as 
\begin{equation}
{\imath}  \partial_t |\Psi\rangle = {\cal H}_{\rm BH} |\Psi\rangle\,.
\label{tdbh}
\end{equation}
For a given $N$, an appropriate many-body basis 
for this bosonic system is the Fock basis, 
$\{ | m_z = (N_1-N_2)/2 \rangle\}$, with 
$m_z=-N/2,\dots, N/2$. A general many-body state, 
$|\Psi\rangle$, can then be written as 
\begin{equation}
|\Psi\rangle = \sum_{m_z=-N/2}^{N/2} c_{m_z} |m_z\rangle \,.
\end{equation}

As customary, the number squeezing parameter is 
defined as~\cite{wine92,esteve08} 
\begin{equation}
\xi_N^2(t)={\Delta \hat{J}_z^2 \over (\Delta \hat{J}_z^2 )_{\rm bin}}\,,
\end{equation}
where 
$\Delta \hat{J}_z^2 \equiv \langle {\hat J}_z^2\rangle 
- \langle {\hat J}_z\rangle^2$ and 
$(\Delta \hat{J}_z^2 )_{\rm bin}=N/4$ for a coherent state 
with $\langle \hat{J}_z\rangle=0$. The many-body state is 
said to be number-squeezed 
if $\xi_N<1$~\cite{kita}. The coherent spin-squeezing 
parameter which accounts for reductions in the 
fluctuations in $\hat{J}_z$, taking 
$\langle \hat{J}_y\rangle=0$, is defined as~\cite{wine92,kita,Sorensen2001}
\begin{equation}
\xi_S^2=  {N (\Delta \hat{J}_z^2) \over \langle \hat{J}_x\rangle^2} =  
{\xi_N^2 \over \alpha^2},
\end{equation}
where the phase coherence of the many-body state is 
$
\alpha(t) = \langle \Psi(t)| 2\hat{J}_x/N|\Psi(t) \rangle\,.$
$\xi_S$ takes into account the delicate compromise 
between improvements in number-squeezing and loss of 
coherence. States with $\xi_S<1$ have been proposed to 
be used in a new Ramsey type atom interferometer with 
an increased phase precision compared to the coherent 
spin state~\cite{wine92}. This gain in precision can be directly 
related to entanglement in the system~\cite{pezze09}.

As detailed in Refs.~\cite{ST08,ours10-2,oursober} a 
systematic expansion of ${\cal H}_N \equiv {\cal H}_{\rm BH}/(N J)$ 
in powers of $h\equiv 1/N$ gives in the semiclassical 
limit, $ N \gg 1$, the pseudo-Schr\"odinger-like equation,
\begin{eqnarray}
{\cal H}_N(z) \psi(z)  
&\equiv&-2 h^2 \partial_z \sqrt{1-z^2}\partial_z \psi(z) + {\cal V}(z) \psi(z)
\label{eq:pse}
\end{eqnarray}
where 
\beq
{\cal V}(z)= -\sqrt{1-z^2}+{1\over2}\Lambda z^2\,,
\eeq
$z=m_z/(N/2)$ , $\Lambda = NU/(2J)$ and $\psi(z) = \sqrt{N/2}\, c_{m_z}$, 
normalized as  $\int_{-1}^1 dz |\psi(z)|^2=1$. The 
corresponding TDSE writes as \beq
i h \partial_t \psi(z,t) = {\cal H}_N
\psi(z,t)
\eeq 
with time measured in units of $1/J$. In these 
units the Rabi time is defined as $t_{\rm Rabi}= \pi/J$. 
Let us emphasize that we construct a systematic expansion on the 
small parameter, $h$, but we do not take the formal limit 
$h\to 0$. The validity of the expansion will improve as 
the considered number of atoms is increased, but it is 
already accurate for $N\gtrsim 50$. 

As explained in~\cite{oursober}, and previously noted 
by other authors~\cite{Jav99,ST08}, for repulsive 
interactions the potential in Fock space, ${\cal V}(z)$, 
is to a very good approximation a harmonic oscillator: 
Neglecting the $z$ dependence of the effective 
mass term and expanding $\sqrt{1-z^2} \simeq 1- z^2/2$ 
in ${\cal V}(z)$, Eq.~(\ref{eq:pse}), reduces to 
\beq
{\cal H}_N \simeq -2h^2 \partial_z^2 +  \frac{1}{8} \omega^2 z^2 \,, 
\label{eq:parab}
\eeq 
with $\omega=2 \sqrt{1+\Lambda}$. 

We have checked that all the expectation values 
of $\hat{J}_i$ and $\hat{J}^2_i$, $i=x, y, z$ computed 
by directly solving the TDSE for the BH Hamiltonian Eq.~(\ref{tdbh}), 
with $\Lambda\equiv \Lambda(t)$ as required by the 
control protocols explained below, and the corresponding 
approximate ones using the solution Eq.~(\ref{eq:pse}) 
together with the explicit expressions given in 
Ref.~\cite{oursober} agree almost perfectly in all calculations 
reported in this article. The only minor discrepancy between the 
results obtained using the continuous version, 
Eq.~(\ref{eq:pse}), and the full TDSE for the Bose-Hubbard Hamiltonian, 
Eq.~(\ref{tdbh}), is explicitly shown in Fig.~\ref{fig3}.

It is worth stressing that Eq.~(\ref{eq:pse}) allows to 
study quantum properties, i.e. squeezing, which are beyond 
the usual fully-classical description~\cite{smerzi97}. 
Setting $h=0$ in Eq.~(\ref{eq:pse}), which removes the 
kinetic term in Fock-space, one obtains Eq.~(5) of 
Ref.~\cite{smerzi97} around $\phi=0$.

\begin{figure}[t]                         
\begin{center}
\includegraphics*[width=8.5cm]{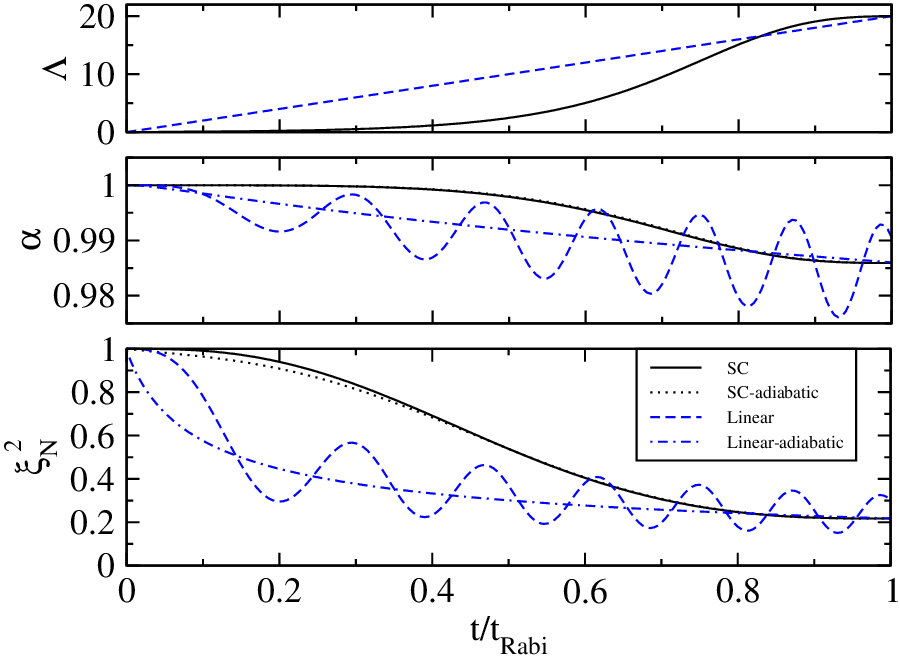}
\caption{(color online) Time evolution of the 
coherence $\alpha(t)$ and the number squeezing 
$\xi_N^2$ of the system obtained solving the 
TDSE for the BH, Eq. (\ref{tdbh}), using the fast adiabatic evolution 
of Eq.~(\ref{eq:sc}) (solid) compared to the case 
of linear ramping of $\Lambda(t)$ 
(dashed). The corresponding instantaneous adiabatic 
results for both the fast adiabatic case and a 
linear ramping are given in dotted and dot-dashed, 
respectively. $\Lambda(0)=0$, $\Lambda(t_{\rm Rabi})=20$, 
and $N=100$.
\label{fig1}}
\end{center}
\end{figure} 

\section{Fast shortcut to adiabatic evolution}
\label{sec2}

Our method for fast adiabatic-like preparation of a given 
ground state requires control of the atom-atom 
interaction at time scales of fractions of the Rabi 
time. In what follows we make use of Eq.~(\ref{eq:parab}) 
to find an optimal solution for the control parameter
$\Lambda$(t) with our shortcut (SC) method (explained below), and 
we later use Eq.~(\ref{tdbh}) with $\Lambda$(t) given 
by the solution found, to compute the squeezing, visibility 
and the other parameters presented in Figs.~\ref{fig1},~\ref{fig2}, 
~\ref{fig3}, and~\ref{fig4}. In the parabolic approximation discussed above 
the protocol developed in Ref.~\cite{muga1} has a 
direct translation into our variables. We need to 
impose the following time variation of 
$\Lambda$~\footnote{Note that the derivations in the 
previous section imply a fixed $J$ and $N$, thus 
the variation in $\Lambda$ has to be due to a 
variation in $U$.}, $\Lambda(t) = \omega^2(t)/4 - 1$, 
where $\omega^2(t)$ obeys the Ermakov equation,  
\beq
\ddot{b}(t)/ J^2 + \omega^2(t) b(t) = \omega_0^2/b^3(t) \ ,
\label{eq:sc}
\eeq 
with time in units of $1/J$ as explained above. 
The key ingredient is to ensure that $b(t)$ 
satisfies the six {\it frictionless conditions}: 
\beqa
b(0)&=&1\,,\nonumber\\ 
b(t_F)&=&r\,, \\
\dot b(0)&=&\ddot b(0)=\dot b(t_F)=\ddot b(t_F)=0 \,,\nonumber
\eeqa 
where
$r=\sqrt{\omega_0/\omega_F}$, 
$\omega_0 = 2 \sqrt{1 + \Lambda_0}$ 
and $\omega_F = 2 \sqrt{1 + \Lambda_F}$. 
The fact that these can be fulfilled 
by an infinite set of $b(t)$, has two 
important consequences: 1) This freedom allows to apply 
tools of Optimal Control Theory (OCT) to produce $b(t)$'s 
which ensure fidelity 1 and satisfy other constraints. 
For instance: a) optimize the value of $t_F$ (in contrast, e.g. 
with the simple ``bang-bang'' methods described in 
the Appendix), b) ensure that 
$\Lambda(t)$ is bound by previously chosen 
experimental values, or c) guarantee that the potential energy 
of the system at intermediate steps is bound by some 
desired value~\cite{muga4,oct}. 2) The method is 
extremely robust: in an experimental realization aiming 
at a final fidelity equal to 1, the success is guaranteed 
provided the frictionless conditions are satisfied, with 
no need to have a precise control during the intermediate 
evolution~\cite{schaff10}. This is an important advantage 
with respect to other protocols, 
e.g.~the ones considered in Ref.~\cite{lapert12}. To 
give an analytic and smooth example we consider here the 
polynomial ansatz for $b(t)$ from Ref.~\cite{muga1}, 
\beq
b(t) = 6(r-1) s^5-15 (r-1) s^4+10 (r-1) s^3+1\,,
\eeq 
with $s= t/t_F$. A brief comparison to another ansatz is presented below.

\section{Results}
\label{sec3}

Our calculations confirm that our method, although 
not exact, is extremely accurate in the many-body 
simulations despite the fact that it was derived for 
the single particle problem in a harmonic oscillator 
Hamiltonian. This is due to the validity of the explained 
mapping of the BH to the single particle pseudo-Schr\"odinger 
equation, which turns out to be well approximated by 
a harmonic oscillator potential. This harmonic approximation 
improves as $N$ is increased because the spread 
of the GS wave function is $\sim 1/\sqrt{N}$, 
thus exploring only the very central part of 
${\cal V}(z)$ in $z\in [-1,1]$. Exceptions 
appear in the ultra-fast preparation as discussed 
later.

\begin{figure}[t]                         
\begin{center}
\includegraphics*[width=8.5cm]{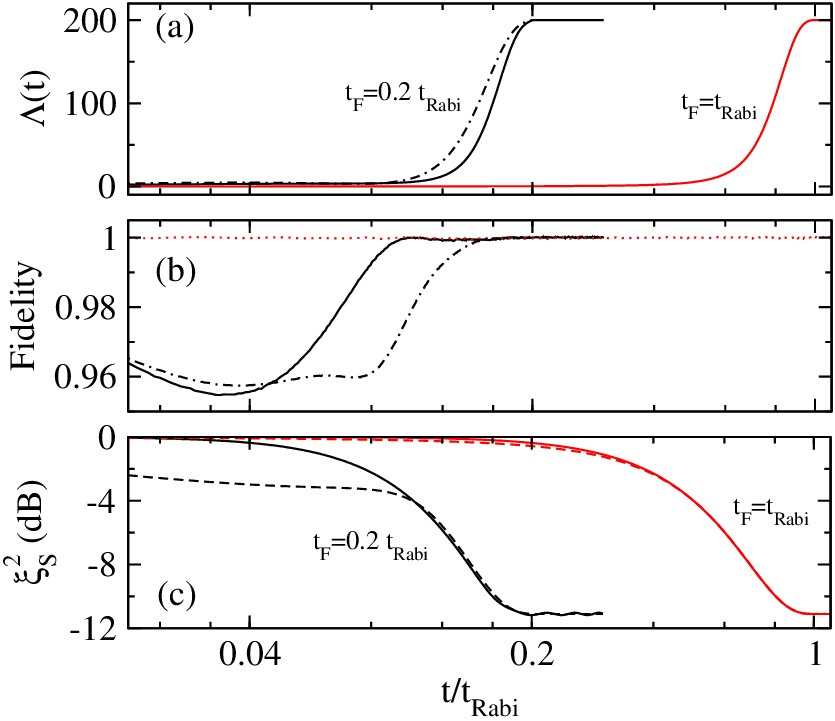}
\caption{(color online) 
(a) Considered $\Lambda(t)$: two final times are considered, 
$t_F=0.2\, t_{\rm Rabi}$, and $1\, t_{\rm Rabi}$. 
(b) Fidelity for the cases considered, note for 
$t_F=t_{\rm Rabi}$ the curve is $\sim 1$ during all the interval. 
The dot-dashed line in panels (a) and (b) corresponds
to a calculation using a non-polynomial ansatz for $b(t)$, see text.
(c) coherent spin 
squeezing $\xi_S^2$ as a function of $t$ using the fast-adiabatic 
process described in the text. The adiabatic $\xi_S^2$, corresponding to 
the instantaneous ground state for each $\Lambda$, is shown in 
dashed lines in the lower panel. The initial and 
final $\Lambda$ are 0 and 200. $N=100$.
\label{fig2}}
\end{center}
\end{figure} 

Let's start by comparing the fast protocol to shortcut the adiabatic following presented above to the 
case of a linear ramping 
{\bf $\Lambda_l(t)=\Lambda(0)+ (t/t_F) (\Lambda(t_F)-\Lambda(0))$. }
In this case we consider a relatively long final 
time of one full Rabi period $t_{\rm Rabi}=\pi/J$, 
see Fig.~\ref{fig1}. For both $\Lambda(t)$  we present the 
results of the corresponding instantaneous ground state 
of the many-body system for comparison. First, we note 
that the SC does produce a final result which has 
the same coherence $\alpha$ and coherent number squeezing 
$\xi_N^2$ as the corresponding adiabatically evolved case. 
In contrast, the linearly ramped system deviates 
notably from the adiabatically evolved state at the 
final time. As we are considering fairly long times, 
the SC is not found to deviate appreciably from 
the instantaneous adiabatic following of the state, 
except for a short period at early times. The transient 
non-adiabaticity will be shown to increase as 
we require shorter $t_F$'s.

The fidelity between the evolved many-body state
$|\Psi(t)\rangle$ and the corresponding instantaneous 
ground-state, $|\langle\Psi(t) | \Psi_{\rm GS}\rangle|$, 
is plotted in Fig~\ref{fig2} and is seen to be extremely 
close to 1 at the final time for both $t_F=$0.2  $t_{\rm Rabi}$ 
and 1 $t_{\rm Rabi}$. 
In the figure we compare the full many-body evolution 
to the corresponding instantaneous ground 
state for different $t_F$. As mentioned above, for 
short times, the fast-adiabatic passage produces 
intermediate many-body states which depart from the 
instantaneous ones, as seen clearly in the drop of 
the fidelity for very short times in the 
$t_F=0.2 \, t_{\rm Rabi}$ case. 

The freedom in choosing $b(t)$ is large, providing a 
very robust protocol when the frictionless conditions 
are satisfied. For example, the use of a non-polynomial 
choice~\footnote{We consider for illustration, 
$b(t)=r^{6 s^5-15 s^4 +10 s^3}$.} for $b(t)$ still produces a fidelity equal to 1 at 
$t=t_F$, see dot-dashed lines in Fig.~\ref{fig2}(a,b). This 
freedom can be exploited e.g. to fulfill actual experimental 
constraints on the control parameters.

\begin{figure}[t]                         
\begin{center}
\includegraphics*[width=8.5cm]{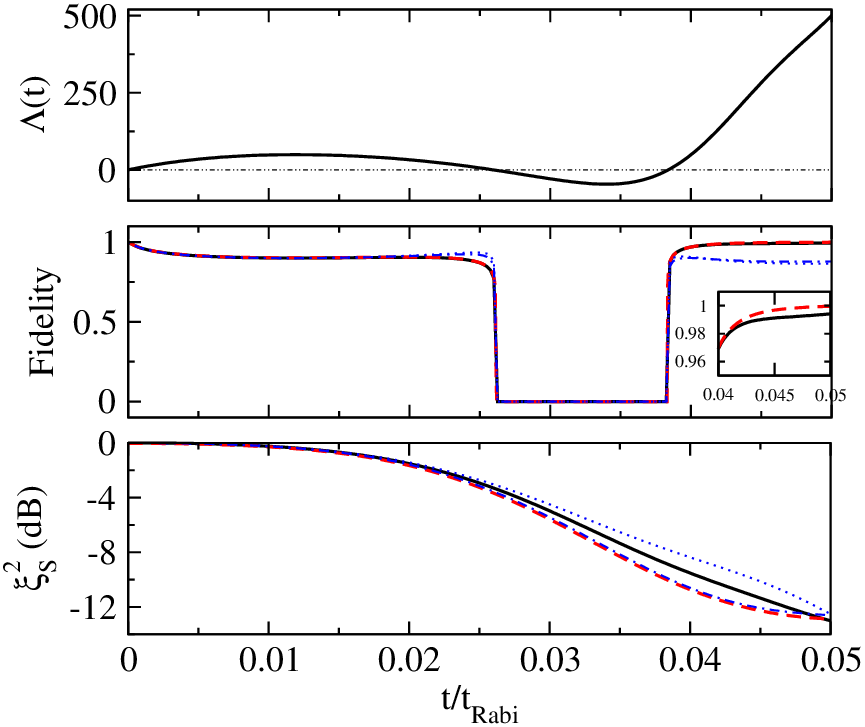}
\caption{(color online) 
Considered $\Lambda(t)$ (upper panel), fidelity (middle 
panel) and coherent spin squeezing $\xi_S^2$ (lower 
panel) as a function of $t$ for the exact Bose-Hubbard 
calculation (solid) and for the Schr\"odinger-like 
equation~(\ref{eq:pse}), (dashed, almost overlapped with the 
solid in the upper and middle panels). The final time $t_F$ is 0.05 
$t_{\rm Rabi}$. The initial and final $\Lambda$ are 0 and 
500. The dotted and dot-dashed lines are obtained assuming 
the actual $\Lambda(t)$ is 10\% smaller or larger than 
the nominal one, respectively. $N=100$. The inset zooms 
on the times close to $t_F$.
\label{fig3}}
\end{center}
\end{figure} 

\subsection{Ultra-short adiabatic production of squeezed states} 

As described in Ref.~\cite{muga1} for a  particle in a 
harmonic oscillator and ultrafast adiabatic evolution, 
the method presented requires a certain period of the 
evolution where the trapping potential is actually reversed 
into an expelling potential (an inverted parabola). 
Through the previously explained mapping, this corresponds 
in our many-body BH problem to a period of time where the 
atom-atom interaction is switched from repulsive to 
attractive, $\Lambda<0$. For attractive interactions the 
potential ${\cal V}(z)$ becomes a double-well potential 
in Fock space and the parabolic approximation of 
Eq.~(\ref{eq:parab}) does only hold for central regions 
in $z$~\cite{oursober}. Thus, one can expect that the 
present method should still work only when the wavefunction 
$\psi(z,t)$ does not spread in $z$ significantly 
during the time where the atom-atom interactions are 
attractive. As seen in Fig.~\ref{fig3} this is indeed 
the case. We consider 
$t_F =0.05\, t_{\rm Rabi}$, $\Lambda(0)=0$ and $\Lambda(t_F)=500$. 
Then the required $\Lambda(t)$ (upper panel) becomes negative 
for $0.025 \lesssim t/t_{\rm Rabi} \lesssim 0.04$. The method 
is seen to work accurately, producing a fidelity $\sim 1$ 
at the final time. The intermediate evolution is 
however highly diabatic, with close to zero fidelity between 
the evolved state and its corresponding adiabatic one, during 
the $\Lambda(t)<0$ evolution. The effect of a systematic 
error in the implementation is analyzed in Fig.~\ref{fig3} 
finding that a 10\% deviation in the value of $\Lambda(t)$ 
results in a decrease of the fidelity to $\sim 0.87$, 
preserving $\xi^2_S\sim -12$dB at $t_F$. 

\subsection{Comparison to existing experimental setups}

To illustrate the improvement in preparation times which 
can be obtained with the protocols discussed above, 
we consider some characteristic parameter values which 
have already been implemented experimentally.

In the experiment of Gross et al.~\cite{gross10}, the 
ultracold atomic cloud is formed by $\sim 400$ atoms of 
$^{87}$Rb. The nonlinear parameter achieved is 
$U= 2 \pi\times 0.063$ Hz, while the linear coupling, 
$2J$, can be varied from 0 Hz to $2 \pi\times 600$ Hz. 
With these conditions, a number squeezing of 
$\xi_N^2\sim -8.2$ dB was achieved after $\sim 20$ ms.

Let us consider a Rabi coupling $2J=2\pi$ Hz 
($t_{\rm Rabi}=1$ s), and the same final value of the 
non-linear term, $U_F=2 \pi\times 0.063$ Hz. Assuming 
an initially non-interacting system $U_i=0$ Hz, and a 
control over $U(t)$ from the initial to the final value 
as required by our protocol, we find that the final value 
of the adiabatic number squeezing is, $\xi_N^2\sim -8.55$ dB. 
To obtain such a highly number-squeezed state in half 
the time taken in the experiment of Gross et al., we 
need $t_F=10$ ms, which corresponds to $t_F=0.005$ $t_{\rm Rabi}$. 
This is achiveable with our protocol requiring an 
ultrafast protocol such as the one presented in Fig.~\ref{fig3}.

In the experiment of Riedel {\it et al.}~\cite{riedel10}., 
the authors trap 1250 atoms with an effective value of 
the nonlinear coupling $U_e=0.49$ Hz. They obtain a best 
value of $\xi_S^2\sim -2.5$ dB after a time of $\sim 15$ ms. 
Using their values, assuming a good control on $U$ from 
$U=0$ Hz to $U=U_e$ and taking a Rabi coupling of 
$2J\sim 2\pi \times 10$ Hz ($t_{\rm Rabi}=0.1$ s) we would 
obtain an adiabatic coherent squeezing of $\xi_S^2\sim -5.1$ dB. 
To produce this squeezing value in half the time used by 
the authors of Ref.~\cite{riedel10}, we need $t_F=7.5$ ms, 
which corresponds to $t_F=0.075$ $t_{\rm Rabi}$. This is similar 
to the ultrafast cases considered in Fig.~\ref{fig3}.

\section{Conclusions}
\label{sec4}

We have presented protocols for fast 
generation of very coherent-spin-squeezed states in 
bosonic Josephson junctions. The attained squeezing 
is the one corresponding to the ordinary adiabatic 
evolution in the case of repulsive atom-atom interactions, 
but requires much shorter preparation times. Ordinary 
adiabatic squeezing is known to improve as $N$ is 
increased as $\xi_S^2 \propto N^{-1}$.
Thus, practical methods of fast-adiabatic driving 
present important advantages for any future experimental 
implementation of BJJs where they are used to produced 
highly squeezed spin states. The present procedures 
require a good control of the time variation of the 
atom-atom scattering length during the desired period, 
a possibility now at hand in current experimental setups 
for BJJ's. The methods have been obtained by 
extending recently developed protocols for fast-adiabatic 
evolution originally devised for a single particle in 
a time dependent harmonic trap, to the Bose-Hubbard 
Hamiltonian. The experimental implementation of the 
proposal would represent a useful step towards the 
fast preparation of many-body entangled quantum resources.

\begin{acknowledgments}
The authors thank M. Oberthaler and T. Zibold for useful 
comments and discussions. This work has been supported 
by FIS2008-01661, 2009-SGR1289, IT472-10, FIS2009-12773-C02-01, 
FIS2008-00784 TOQATA and the UPV/EHU under program UFI 11/55. 
B.~J.-D. is supported by the Ram\'on y Cajal program. E. T. 
acknowledges financial support from the Basque Government 
(Grants No. BFI08.151).
\end{acknowledgments}

\appendix

\section{Bang-bang methods}
\label{app:bang}

\begin{figure}[t]                         
\begin{center}
\includegraphics*[width=8.cm]{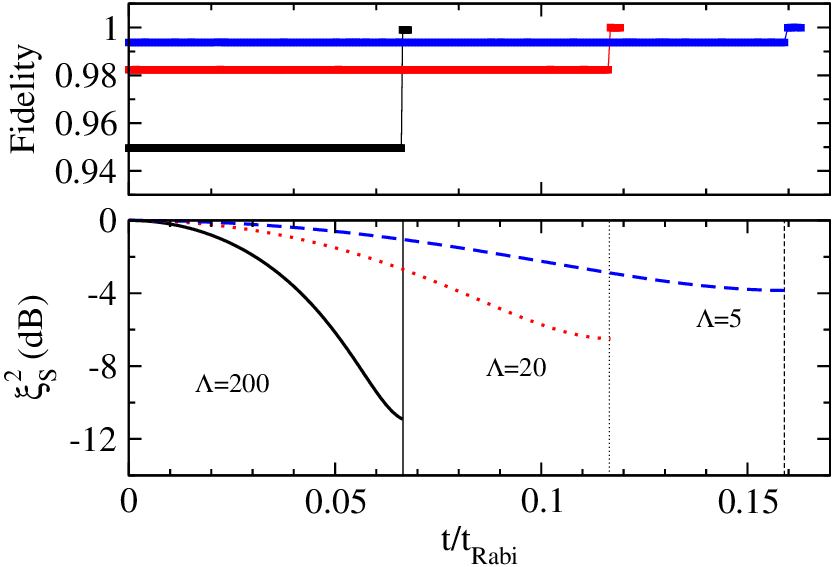}
\caption{(color online) Coherent spin squeezing attainable 
using a simple ``Bang-bang'' method as described in the 
text (lower panel). The fidelity between the evolved state 
and its corresponding instantaneous evolution is given in 
the upper panel. $\Lambda(0)=0$ in all cases, while 
$\Lambda(t_{\rm bang})=5$ (dashed), 20 (dotted), and 
200 (solid). $N=100$.
\label{fig4}}
\end{center}
\end{figure} 

Another set of protocols known to produce 
fast-adiabatic evolution in harmonic oscillator 
Hamiltonians are the so-called ``bang-bang'' 
methods~\cite{kosloff}. The translation of one of the simplest 
versions of these methods to the present problem 
requires the following steps assuming that we 
start as before from an initial ground state 
corresponding to $\Lambda(0)=\Lambda_0$, and want 
to end up in the ground state corresponding to 
$\Lambda_F$: a) at $t=0$ set 
$\Lambda= \Lambda_{\rm B}=\omega_0 \omega_F/4-1$ 
(where $\omega_0=\sqrt{1+\Lambda_0}$ and 
$\omega_F=\sqrt{1+\Lambda_F}$), b) let the system 
evolve at fixed $\Lambda_{\rm B}$ 
until the time 
$t=t_{\rm B} = t_{\rm Rabi}/(2\sqrt{\omega_0 \omega_F})$, 
and c) change $\Lambda_{\rm B}$ to $\Lambda_F$. 
Note that the final time is not arbitrary, as before, but is 
fixed by the initial and final values of $\Lambda$. 
This makes it in practice more unstable
with respect to small errors~\cite{muga6}. 
Under ideal conditions (perfect timing) the 
method is seen to work well producing fairly 
low values of $\xi_S^2$ at times of the order of 
$t_{\rm Rabi}/10$ for $N =100$ atoms with a 
fidelity of $\sim 1$ at the final time, see 
Fig.~\ref{fig4}.


\end{document}